\documentclass{article}

\usepackage[final]{ml4mols_2020}

\usepackage[utf8]{inputenc} 
\usepackage[T1]{fontenc}    
\usepackage{hyperref}       
\usepackage{url}            
\usepackage{booktabs}       
\usepackage{amsfonts}       
\usepackage{nicefrac}       
\usepackage{microtype}      
\usepackage{alphabeta}
\usepackage{amsmath,amsfonts,amssymb}
\usepackage{graphicx}
\usepackage{float}
\usepackage{wrapfig}
\usepackage{caption}
\usepackage{subcaption}
\usepackage{lipsum}  
\usepackage{amsmath}
\usepackage{authblk}
\usepackage{color}
\usepackage{wrapfig}
\usepackage{rotating}
\setcitestyle{numbers}
\title{Non-equilibrium molecular geometries in graph neural networks}

\author[1]{\textbf{Ali Raza}\thanks{\texttt{razaa@oregonstate.edu}}~~}
\author[2]{\textbf{E. Adrian Henle}~}
\author[1]{\textbf{Xiaoli Fern} \thanks{\texttt{xfern@oregonstate.edu}}~~}
\affil[1]{School of Electrical Engineering and Computer Science, Oregon State University}
\affil[2]{School of Chemical, Biological, and Environmental Engineering, 
   Oregon State University}

\begin{document}

\maketitle

\begin{abstract}

Graph neural networks have become a powerful framework for learning complex structure-property relationships and fast screening of chemical compounds. Recently proposed methods have demonstrated that using 3D geometry information of the molecule along with the bonding structure can lead to more accurate prediction on a wide range of properties. 
A common practice is to use 3D geometries computed through density functional theory (DFT) for both training and testing of models. However, the computational time needed for DFT calculations can be prohibitively large. Moreover, many of the properties that we aim to predict can often be obtained with little or no overhead on top of the DFT calculations used to produce the 3D geometry information, voiding the need for a predictive model. To be practically useful for high-throughput chemical screening and drug discovery, it is desirable to work with 3D geometries obtained using less-accurate but much more efficient non-DFT methods.
In this work we investigate the impact of using non-DFT conformations in the training and the testing of existing models and propose a data augmentation method for improving the prediction accuracy of classical forcefield-derived geometries. 
\end{abstract}

\section{Introduction}
\label{sec:introduction}
Computational screening can accelerate drug discovery by efficiently exploring a much larger space of compounds and identifying promising candidates for experimental testing. In recent years machine learning (ML) has achieved great success in the fast screening of chemical structures. It is more efficient than physics-based methods and more accurate than traditional rule-based methods~\cite{stokes2020deep}\cite{gomez2016design}\cite{zhavoronkov2019deep}. Graph neural networks (GNNs) have been considered as the ML method of choice for such problems because they conveniently take molecular graphs as input to learn continuous embeddings, without the need for feature engineering. Recent developments in GNNs have incorporated geometry information. 
It is a common practice for such methods to report their performance by training and testing the proposed models using optimized equilibrium 3D molecular geometries. Obtaining minimum-energy atomic coordinates, e.g., using density functional theory (DFT), for a new molecule is a computationally intensive task and impractical for fast chemical screening. Even if the time required is not critical, and computationally expensive procedures are acceptable in exchange for better accuracy, the properties we are interested in predicting can often be obtained with little or no overhead on top of the computation needed for obtaining the 3D geometry.
In such cases, it is preferable to directly obtain these properties using DFT as opposed to using machine learning, which uses DFT computed geometries to make relatively less accurate predictions. As such, we argue that, to be practically useful, ML models should work with geometries obtained using methods that do not require computationally expensive operations like DFT. This, however, is not the common practice in the literature. In this work, we examine the impact of using non-equilibrium geometry with state-of-the-art GNN models, trained using DFT or non-DFT geometries. Furthermore, we propose a simple scheme of data augmentation to make the models more robust and improve the prediction accuracy using non-equilibrium geometries computed by efficient non-DFT methods.

\section{Background}
\label{sec:background}
\paragraph{Graph Neural Networks.} 

Graph neural networks (GNNs) have been developed to process graph data~\cite{kipf2016semi}\cite{velivckovic2017graph}, such as the graph representations of molecular structures (with atoms as nodes and chemical bonds as edges). 
GNNs follow a message passing (recursive neighborhood aggregation) scheme by embedding each node in a high-dimensional space and aggregating feature vectors (node and/or edge embedding) of its neighbors to compute its new feature vector~\cite{gilmer2017neural}. After $T$ iterations, each node has structural information of its $T$-hop neighborhood (or local bonding environment). 
Original GNNs were defined for regular graph structures (2D graphs). Recently, new GNN architectures have been proposed to use 3D molecular graphs, which include the 3D coordinates for each node in addition to the bonding information. These models incorporate the distances and angular information derived from the 3D geometry of the graph to enhance the message passing between neighbors. See Table~\ref{table:gnn3d} for a summary of these works. 

\begin{table}[!b]
\centering
\begin{tabular}{lp{0.42\linewidth}l}
\toprule
Name      & Approach                                                                                                                      & Dataset used                    \\
\cmidrule(r){1-3}
SchNet~\cite{schutt2018schnet}    &  distance information using continuous-ﬁlter convolutional layers & QM9~\cite{ramakrishnan2014quantum}, MD17~\cite{chmiela2017machine}, ISO17~\cite{schutt2017schnet}                \\
PhysNet~\cite{unke2019physnet}   & integrates both the node features and distance information in the proposed interaction block                                  & QM9, MD17, ISO17, SN2~\cite{unke2019physnet} \\
DimeNet~\cite{klicpera2020directional}   & include distance and angle information in the interaction block                                                               & QM9, MD17                       \\
SphereNet~\cite{liu2021spherical} & use radial distance, polar angle, and the azimuthal angle                                                                     & QM9, Open Catalyst~\cite{chanussot2010open}, MD17       \\
\bottomrule \\
\end{tabular}
\caption{Recently proposed GNNs for 3D graphs that use molecular geometry for more accurate predictions.}
\label{table:gnn3d}
\end{table}
\paragraph{3D-conformation of molecules. }
Molecules are intuitively represented as 3D graphs.
For molecules with flexible/rotatable bonds, there may exist multiple locally-optimal spatial configurations, known as conformations.
Molecular conformations may be determined experimentally; this, however, is time-consuming and requires special instrumentation.
The prevailing technique for calculating accurate conformations is density functional theory (DFT), which seeks a numerical solution to the Schrödinger equation under the Born-Oppenheimer approximation.
The trade-off for the accuracy of DFT methods is computational expense, with time complexity $O(e^3)$ where $e$ is the number of electrons in the system.  For larger molecules or molecules containing heavy atoms, runtime becomes impractical for high-throughput screening.
Classical methods for obtaining conformations minimize, with respect to atomic coordinates, a force-field energy function based on, e.g. the Lennard-Jones interaction potential.
Although these classical methods provide relatively cruder approximations of molecular structure \cite{kanal2018sobering}, their efficiency renders them highly practical for generation of conformer ensembles for use in rapid computational screening of novel structures.

\section{Impact of conformations}
\label{sec:impact_conformations}

\begin{table}[]
\centering
\begin{tabular}{llccccccc}
\toprule
                    &  &    \multicolumn{7}{c}{Target (Unit)}    \\ 
 \cmidrule(r){3-9}
                         &                            & $\mu$   & $\alpha$   & $\epsilon_{\textrm{HOMO}}$    & $\epsilon_{\textrm{LUMO}}$    & $\Delta \epsilon$       & $\langle R^2 \rangle$    & $\textrm{ZPVE}$    \\ 
Model                         &        Method                    &  (D)   & ($a_0^3$)   &  (meV)   &  (meV)   &  (meV)      &  ($a_o^2$)   &  (meV)   \\ 
 \cmidrule(r){1-9}                        
GCCN                     & B/B                        & 0.621 & 0.947 & 269     & 404   & 615    & 100.7    & 26.04 \\
\\
SchNet                  & B/B                         & 0.923 & 1.737 & 393     & 727   & 801    & 154.3    & 36.83  \\
                         & DFT/DFT                    & 0.033 & 0.235 & 41      & 34    & 63     & 0.073    & 1.7 \\
                         & DFT/OB                     & 0.875 & 0.723 & 131     & 164   & 200    & 49.7     & 12.21 \\
                         & OB/OB                      & 0.477 & 0.364 & 99      & 96    & 119    & 19.5     & 4.23 \\
\\ 
DimeNet                  & B/B                        & 0.676 & 0.814 & 254     & 576   & 600    & 30.5     & 22.9 \\
                         & DFT/DFT                    & 0.029 & 0.047 & 27.8    & 19.7  & 34.8   & 0.331    & 1.29 \\
                         & DFT/OB                     & 0.718 & 1.08  & 169     & 206   & 197    & 45.9     & 50 \\
                         & OB/OB                      & 0.351 & 0.197 & 72      & 69.4  & 97.5   & 15.9     & 3.6 \\ 
\bottomrule \\
\end{tabular}
\caption{Quantitative results}
\label{table:results}
\end{table}
It is common practice to use DFT conformations to train GNN models; however, with DFT being computationally expensive,  we are interested in understanding how well different models can work when applied with 3D conformations obtained using more efficient but less optimal non-DFT methods. In other words, do we see similar performance benefit from using distances/angular information when using less accurate conformations?
Specifically, we examine two types of models: models trained on DFT conformations and models trained on non-DFT conformations; and test their performances in a practically relevant scenario where only non-DFT conformations are available during testing. 

\paragraph{GNN models used.} We experiment with recently proposed SchNet~\cite{schutt2018schnet} and DimeNet~\cite{klicpera2020directional} models. SchNet incorporates relative distances based on atomic positions. DimeNet embeds atoms via a set of messages (i.e., edge embeddings) and leverages the directional information by transforming messages based on the angle between them. 
To compare these models to those that only consider the bond graph, we also create a modified version of both SchNet and DimeNet to keep their respective architecture but only use the bonding information and compare them with Gated Graph Sequence Neural Networks (GGCN)~\cite{li2015gated}, which does not consider 3D geometries of the molecules.

\paragraph{Dataset.} We use the QM9 computational dataset, a widely used benchmark dataset for molecular property prediction consisting of density functional theory (DFT)-optimized 3D coordinates and energies~\cite{ramakrishnan2014quantum}. QM9 contains geometric, energetic, electronic, and thermodynamics properties for a subset of GDB-17 structures~\cite{ruddigkeit2012enumeration} comprising up to 9 non-hydrogen atoms. 
All physical properties are computed with the B3LYP functional and 6-31G(2df,p) basis set \cite{ramakrishnan2014quantum}.The dataset is split into three sets, where the training set contains $83,092$, the validation set contains $10,386$, and the test set contains $10,387$ molecules. 
We use Open Babel (OB)~\cite{o2011open} to generate non-DFT conformations for the QM9 dataset.  Open Babel is a chemical toolbox that offers support for molecular mechanics and provides the ability to generate a reasonable 3D structure given the connectivity (bonding) information by following a series of steps~\cite{yoshikawa2019fast}. First, it uses a combination of rules and fragment templates to generate a 3D structure. This 3D structure is subsequently cleaned (conformer searching) without altering the stereochemistry. Conformer searching algorithms adopt a torsion-driving approach by setting torsion angles to one of the allowed values. Initial 3D structure is cleaned by a) steepest descent geometry optimization with the MMFF94 forcefield, b) weighted rotor conformational search and c) conjugate gradient geometry optimization. All of these steps ensure that the generated conformation is likely to be the minimum energy conformer.

\paragraph{Results.} Quantitative results are shown in Table~\ref{table:results}. Results for training and testing using DFT conformations (DFT/DFT) for SchNet and DimeNet are provided by the original papers. When we test on the OB conformations with models trained on DFT conformations (DFT/OB), not surprisingly there is a significant loss of accuracy for all targets and models. 
Interestingly, DimeNet appears to be more severely affected compared to SchNet, possibly due to its complex architecture being more rigid and prone to overfitting to the DFT conformations. Nonetheless, in most cases, DFT/OB still performs better than completely ignoring the geometry information and using only the bonding graphs (B/B). 

When models are both trained and tested on OB conformations (OB/OB), we obtain better performance compared to DFT/OB and also DimeNet performs better than SchNet. The OB/OB results show that in practice, where a new chemical compound with non-DFT conformations is considered for fast screening, the deployed models should be trained using non-DFT conformations. Furthermore, the community should shift their practice and report results of using non-DFT conformations to reflect what to expect in practice.

\section{Data augmentation for improved robustness/generalization}\label{sec:data_augmentation}
The results in Section~\ref{sec:impact_conformations} show that when using 3D geometry information in modeling, it is preferable to train your model using conformations that you expect to test on. Unfortunately, this implies that we will need to retrain our model if a different method is used for generating the conformation in testing. Can we remove this need for re-training by making training more robust?  Toward this goal, we propose a simple strategy based on data augmentation that can reduce the sensitivity of the trained model to the specific conformations used and improve robustness and generalization. 

The key idea is to augment each training example with multiple low-energy conformations to increase the diversity of the training and reduce the sensitivity to the variations in conformations. 
Each conformation is a plausible spatial arrangement at a local energy minimum, collectively representing an ensemble of observable states (i.e. likely model inputs) for the molecule.
In principle, this augmentation expected to result in decreased dependence of the model on exact geometry information.  
Additionally, the model should learn to generate more consistent predictions for ground-state molecular properties when conformations other than the global energy minimum are given as input.

We use confab~\cite{o2011confab} to generate multiple low-energy conformers for each QM9 molecule. Confab uses the torsion-driving approach to generate conformations. It iterates through the set of allowed torsion angles for each rotatable bond. Energy is evaluated using the MMFF94 forcefield. Diversity is measured using the heavy-atom root-mean-square deviation (RMSD) relative to the already stored conformers. We use the DFT conformation as the initial conformation for confab. We use $50.0$ kcal/mol as energy cutoff and $0.5$ as RMSD cutoff. Depending on the structure's rotatable bonds, confab generates $1$ to $400$ conformations for each QM9 molecule. We randomly select $30$ conformations at most for each molecule. Each conformation of a molecule will serve as a different graph datum (training example) with the same target value.  Data augmentation results in an increase of training data size from $83,092$ to $519,655$ graphs. 

Results of using data augmentation (DA/OB) to effectively train models are shown in Table~\ref{table:results}. SchNet-DA/OB performs better than OB/OB on 4 properties, and DimeNet-DA/OB achieves better accuracy than OB/OB on all 7 properties. DimeNet is a complex model with many layers. It uses both distances between nodes and angles between edges. This might be the reason why DimeNet is more sensitive to conformations being used for training/testing as well as benefits more from the  augmentation than SchNet which only uses distance information.

\begin{table}[]
\centering
\begin{tabular}{llccccccc}
\toprule
                    &  &    \multicolumn{7}{c}{Target (Unit)}     \\ 
 \cmidrule(r){3-9}
                         &                            & $\mu$   & $\alpha$   & $\epsilon_{\textrm{HOMO}}$    & $\epsilon_{\textrm{LUMO}}$    & $\Delta \epsilon$       & $\langle R^2 \rangle$    & $\textrm{ZPVE}$    \\ 
Model                         &        Method                    &  (D)   & ($a_0^3$)   &  (meV)   &  (meV)   &  (meV)      &  ($a_o^2$)   &  (meV)   \\ 
 \cmidrule(r){1-9}                        
SchNet                  & OB/OB                        &   0.477   &   0.364   &   99      &   96      &   119     &   19.5     &   4.23 \\
                        & DA/OB                        &   0.459   &   0.259   &   92.9    &   89      &   125     &   20.68    &   4.45 \\
\\
DimeNet                 & OB/OB                        &   0.351   &   0.197   &   72      &   69.4    &   97.5    &   15.9     &   3.6 \\ 
                        & DA/OB                        &   0.337   &   0.176   &   63.8    &   58.46   &   91.9    &   14.6    &   3.2  \\
\bottomrule \\
\end{tabular}
\caption{Using data augmentation (DA) to improve prediction accuracy for non-DFT conformations}
\label{table:results_da}
\end{table}
\section{Conclusion and discussion}
\label{sec:conclusion}
It is a common practice in the literature to use DFT-optimized 3D geometries in GNNs. Machine learning models, to be practically useful for fast screening of chemical compounds, should work with geometries not obtained using computationally expensive methods like DFT. We show that there is a significant loss of accuracy when new chemical compounds with non-DFT conformations are tested with models trained on only DFT-conformations. 
Deployed models should be trained and tested using geometries from the same level of theory as what is expected in the predictive application.
Additionally, we propose a data augmentation scheme to reduce the sensitivity of the trained models to the specific conformations and improve robustness and generalization.

\begin{ack}
The authors acknowledge the National Science Foundation for support under grants No.\ 1920945 and No.\ 1521687.
\end{ack}

\bibliographystyle{unsrt}
\bibliography{references}

\begin{thebibliography}{10}

\bibitem{stokes2020deep}
Jonathan~M Stokes, Kevin Yang, Kyle Swanson, Wengong Jin, Andres Cubillos-Ruiz,
  Nina~M Donghia, Craig~R MacNair, Shawn French, Lindsey~A Carfrae, Zohar
  Bloom-Ackermann, et~al.
\newblock A deep learning approach to antibiotic discovery.
\newblock {\em Cell}, 180(4):688--702, 2020.

\bibitem{gomez2016design}
Rafael G{\'o}mez-Bombarelli, Jorge Aguilera-Iparraguirre, Timothy~D Hirzel,
  David Duvenaud, Dougal Maclaurin, Martin~A Blood-Forsythe, Hyun~Sik Chae,
  Markus Einzinger, Dong-Gwang Ha, Tony Wu, et~al.
\newblock Design of efficient molecular organic light-emitting diodes by a
  high-throughput virtual screening and experimental approach.
\newblock {\em Nature materials}, 15(10):1120--1127, 2016.

\bibitem{zhavoronkov2019deep}
Alex Zhavoronkov, Yan~A Ivanenkov, Alex Aliper, Mark~S Veselov, Vladimir~A
  Aladinskiy, Anastasiya~V Aladinskaya, Victor~A Terentiev, Daniil~A
  Polykovskiy, Maksim~D Kuznetsov, Arip Asadulaev, et~al.
\newblock Deep learning enables rapid identification of potent ddr1 kinase
  inhibitors.
\newblock {\em Nature biotechnology}, 37(9):1038--1040, 2019.

\bibitem{kipf2016semi}
Thomas~N Kipf and Max Welling.
\newblock Semi-supervised classification with graph convolutional networks.
\newblock {\em arXiv preprint arXiv:1609.02907}, 2016.

\bibitem{velivckovic2017graph}
Petar Veli{\v{c}}kovi{\'c}, Guillem Cucurull, Arantxa Casanova, Adriana Romero,
  Pietro Lio, and Yoshua Bengio.
\newblock Graph attention networks.
\newblock {\em arXiv preprint arXiv:1710.10903}, 2017.

\bibitem{gilmer2017neural}
Justin Gilmer, Samuel~S Schoenholz, Patrick~F Riley, Oriol Vinyals, and
  George~E Dahl.
\newblock Neural message passing for quantum chemistry.
\newblock In {\em International conference on machine learning}, pages
  1263--1272. PMLR, 2017.

\bibitem{schutt2018schnet}
Kristof~T Sch{\"u}tt, Huziel~E Sauceda, P-J Kindermans, Alexandre Tkatchenko,
  and K-R M{\"u}ller.
\newblock Schnet--a deep learning architecture for molecules and materials.
\newblock {\em The Journal of Chemical Physics}, 148(24):241722, 2018.

\bibitem{ramakrishnan2014quantum}
Raghunathan Ramakrishnan, Pavlo~O Dral, Matthias Rupp, and O~Anatole
  Von~Lilienfeld.
\newblock Quantum chemistry structures and properties of 134 kilo molecules.
\newblock {\em Scientific data}, 1(1):1--7, 2014.

\bibitem{chmiela2017machine}
Stefan Chmiela, Alexandre Tkatchenko, Huziel~E Sauceda, Igor Poltavsky,
  Kristof~T Sch{\"u}tt, and Klaus-Robert M{\"u}ller.
\newblock Machine learning of accurate energy-conserving molecular force
  fields.
\newblock {\em Science advances}, 3(5):e1603015, 2017.

\bibitem{schutt2017schnet}
Kristof~T Sch{\"u}tt, Pieter-Jan Kindermans, Huziel~E Sauceda, Stefan Chmiela,
  Alexandre Tkatchenko, and Klaus-Robert M{\"u}ller.
\newblock Schnet: A continuous-filter convolutional neural network for modeling
  quantum interactions.
\newblock {\em arXiv preprint arXiv:1706.08566}, 2017.

\bibitem{unke2019physnet}
Oliver~T Unke and Markus Meuwly.
\newblock Physnet: a neural network for predicting energies, forces, dipole
  moments, and partial charges.
\newblock {\em Journal of chemical theory and computation}, 15(6):3678--3693,
  2019.

\bibitem{klicpera2020directional}
Johannes Klicpera, Janek Gro{\ss}, and Stephan G{\"u}nnemann.
\newblock Directional message passing for molecular graphs.
\newblock {\em arXiv preprint arXiv:2003.03123}, 2020.

\bibitem{liu2021spherical}
Yi~Liu, Limei Wang, Meng Liu, Xuan Zhang, Bora Oztekin, and Shuiwang Ji.
\newblock Spherical message passing for 3d graph networks.
\newblock {\em arXiv preprint arXiv:2102.05013}, 2021.

\bibitem{chanussot2010open}
L~Chanussot, A~Das, S~Goyal, T~Lavril, M~Shuaibi, M~Riviere, K~Tran,
  J~Heras-Domingo, C~Ho, W~Hu, et~al.
\newblock The open catalyst 2020 (oc20) dataset and community challenges. arxiv
  e-prints 2020.
\newblock {\em arXiv preprint arXiv:2010.09990}, 2010.

\bibitem{kanal2018sobering}
Ilana~Y Kanal, John~A Keith, and Geoffrey~R Hutchison.
\newblock A sobering assessment of small-molecule force field methods for low
  energy conformer predictions.
\newblock {\em International Journal of Quantum Chemistry}, 118(5):e25512,
  2018.

\bibitem{li2015gated}
Yujia Li, Daniel Tarlow, Marc Brockschmidt, and Richard Zemel.
\newblock Gated graph sequence neural networks.
\newblock {\em arXiv preprint arXiv:1511.05493}, 2015.

\bibitem{ruddigkeit2012enumeration}
Lars Ruddigkeit, Ruud Van~Deursen, Lorenz~C Blum, and Jean-Louis Reymond.
\newblock Enumeration of 166 billion organic small molecules in the chemical
  universe database gdb-17.
\newblock {\em Journal of chemical information and modeling},
  52(11):2864--2875, 2012.

\bibitem{o2011open}
Noel~M O'Boyle, Michael Banck, Craig~A James, Chris Morley, Tim Vandermeersch,
  and Geoffrey~R Hutchison.
\newblock Open babel: An open chemical toolbox.
\newblock {\em Journal of cheminformatics}, 3(1):1--14, 2011.

\bibitem{yoshikawa2019fast}
Naruki Yoshikawa and Geoffrey~R Hutchison.
\newblock Fast, efficient fragment-based coordinate generation for open babel.
\newblock {\em Journal of cheminformatics}, 11(1):1--9, 2019.

\bibitem{o2011confab}
Noel~M O'Boyle, Tim Vandermeersch, Christopher~J Flynn, Anita~R Maguire, and
  Geoffrey~R Hutchison.
\newblock Confab-systematic generation of diverse low-energy conformers.
\newblock {\em Journal of cheminformatics}, 3(1):1--9, 2011.

\end{thebibliography}

\end{document}